\documentclass{aa}  
\usepackage{graphicx}
\usepackage{txfonts}
\usepackage{xspace}

\usepackage[switch]{lineno}

\newcommand{\rxj} {RX J1713.7$-$3946\xspace}
\newcommand {\xmm} {{\it XMM-Newton\xspace} }

\newcommand {\chandra} {{\it Chandra}\xspace}

\newcommand {\g} {$\gamma$\xspace}

\begin{document}

   \title{Measurement of the X-ray proper motion in the south-east rim of RX J1713.7$-$3946}
  \author{Fabio Acero\inst{1}, Satoru Katsuda\inst{2}, Jean Ballet\inst{1}, \and Robert Petre\inst{3}}
  \titlerunning{X-ray proper motion in \rxj}
  \authorrunning{Acero et al.}

 \institute{ \footnotesize{Laboratoire AIM, IRFU/SAp - CEA/DRF - CNRS - Universit\'e Paris Diderot, Bat. 709, CEA-Saclay, Gif-sur-Yvette Cedex, France }\email{fabio.acero@cea.fr} \\
                \and
                 \footnotesize{ Department of Physics, Faculty of Science \& Engineering, Chuo University, 1-13-27 Kasuga, Bunkyo,Tokyo 112-8551, Japan} \\
                 \and
                 \footnotesize{NASA Goddard Space Flight Center, Greenbelt, MD 20771 }}

   \date{Received 30 August 2016; accepted 18 November 2016}

 
  \abstract{
We report on the first proper motion measurement in the supernova remnant RX J1713.7$-$3946 using the \textit{XMM-Newton} X-ray telescope on a 13 year time interval.  
This expansion measurement is carried out in the south-east region of the remnant, where two sharp filament structures are observed. 
For the outermost filament, the proper motion is  $0.75^{+0.05}_{-0.06} \pm 0.069_{\rm syst}$  arcsec yr$^{-1}$  which is equivalent
 to a shock speed of $\sim$3\,500 km s$^{-1}$ at a distance of 1 kpc. 
 In contrast with the bright north-west region, where the shock is interacting with the border of the cavity,
  the shock in the south-east region is probably expanding in the original ambient medium carved by the progenitor and can be used to 
  derive the current density at the shock and the age of the remnant.
  In the case where the shock is evolving in a wind profile ($\rho \propto r^{-s}$, $s=2$) or in a uniform medium ($s=0$), 
  we estimate an age of $\sim$2\,300 yrs and $\sim$1\,800 yrs respectively for an ejecta power-law index of $n=9$. 
  The specific case of an ejecta power-law index of $n=7$, and $s=0$, yields an age of $\sim$1\,500 yrs,
   which would reconcile RX J1713.7$-$3946 with the historical records of SN 393.
In all scenarios, we derive similar upstream densities of the order of 0.01 cm$^{-3}$, compatible with the lack of thermal X-rays from the shocked 
ambient medium.  }

\keywords{ISM: individual (\rxj) --- ISM: supernova remnants ---  X-rays: ISM}

   \maketitle
%

\section{Introduction} \label{sec:intro}

Shell-like X-ray synchrotron emission in supernova remnants (SNRs) is the evidence 
that electrons can be accelerated to very high energies (up to 100 TeV) at the shock.
All such SNRs are young ($t<5\,000$ yrs) and exhibit  high shock speeds ($V_{\rm sh}>2\,000$ km s$^{-1}$).
These are ideal conditions in which diffusive shock acceleration 
can take place. Therefore, these sources provide 
the best targets to study the detailed physics of cosmic ray (CR) acceleration in SNRs.

While very young historical SNRs,  e.g. Tycho, Kepler, SN 1006 typically show a mix of 
both synchrotron and thermal emission (from ejecta and shocked ambient medium), some slightly older SNRs, e.g. \rxj, Vela Jr and HESS J1731-347,
exhibit spectra dominated by X-ray synchrotron emission. Despite a decade of search, no thermal X-ray emission was detected in any of these SNRs
until \citet{katsuda15}  recently reported the first detection of thermal X-rays from the central region of \rxj, which  probably originates from shocked ejecta, rather than shocked interstellar medium (ISM). 
The measured  abundance ratio suggest that the progenitor of \rxj was a relatively low-mass star ($<$ 20 M$_{\sun}$), consistent with 
the estimate of 12-16 M$_{\sun}$ \citep{cassam04} based on the effect of stellar winds of the progenitor star on the surrounding medium.

\begin{figure*}[t]

\includegraphics[bb=30 200 565 590,clip,width=8.7cm]{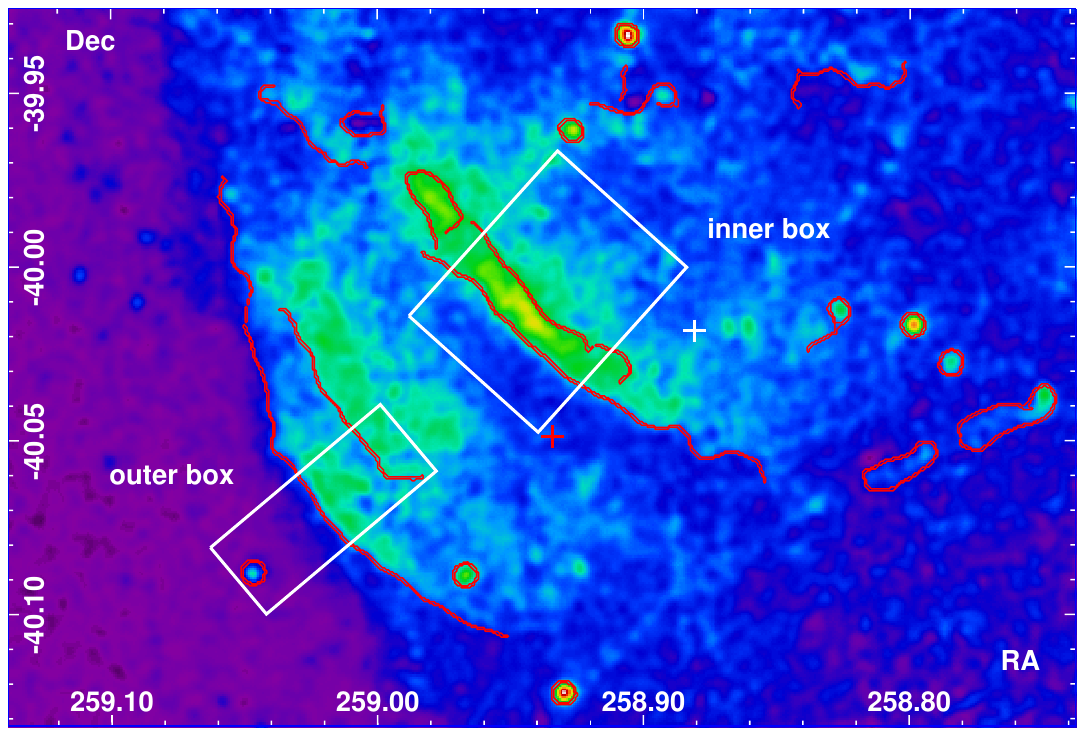} \hspace{9mm}
\includegraphics[bb=40 200 565 590,clip,width=8.7cm]{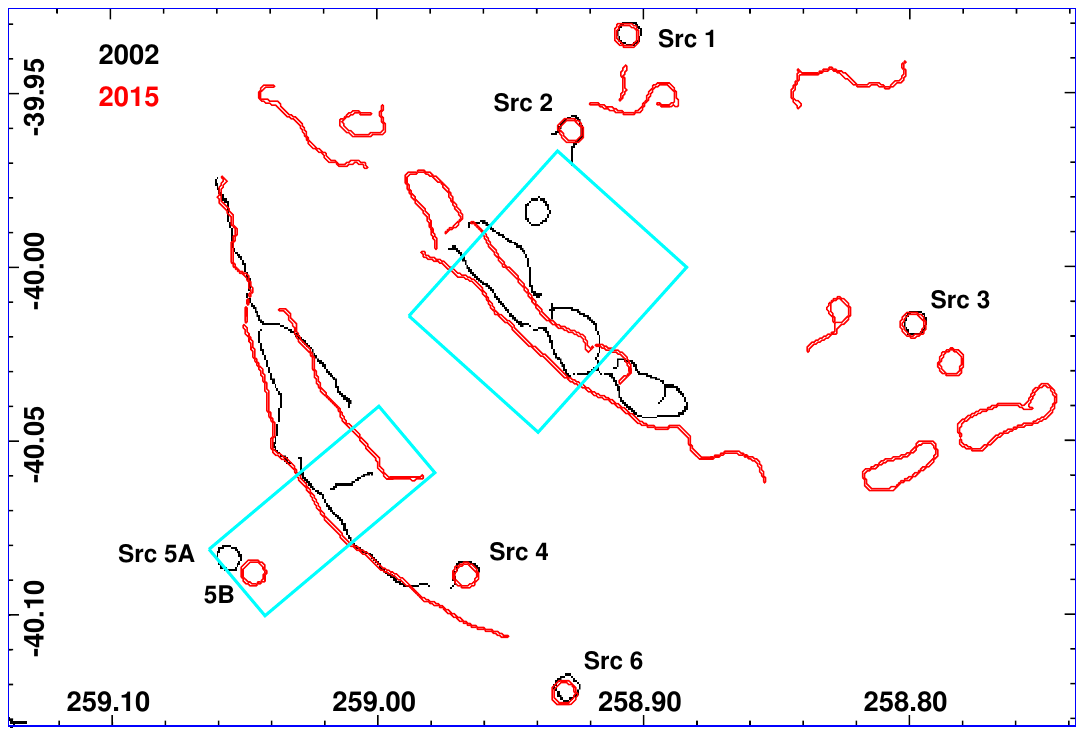}

\caption{ \textit{Left panel}: exposure corrected and background subtracted count map in the 0.6-6 keV energy band of the 2015  \xmm observation. The image is smoothed with a Gaussian of 20 arcsec kernel radius. The red contours are the results of the Canny edge detection algorithm discussed in the main text.  The white boxes represent the regions used to extract the radial profiles shown in Figure \ref{profiles} and the white and red crosses the 2002 and 2015 pointing positions. \textit{Right panel}: edges detected in the 2002 (black) and 2015 (red) observations. The point sources discussed in the image registration process are labeled Src 1 to Src 6. }
\label{XMMimage}
\end{figure*}

\rxj  is the brightest SNR at TeV energies and is considered the prototypical CR accelerator. Despite clear evidence of efficient electron acceleration from X-ray synchrotron, it is unclear whether \g-rays originate from accelerated hadrons (dominating the Galactic CR composition) or electrons. GeV observations with {\it Fermi}-LAT show a hard spectrum \citep[$\Gamma=1.50\pm0.11$,][]{abdo11} that is incompatible with purely hadronic \g-rays stemming from a canonical $E^{-2}$ proton population, but can be  represented by inverse Compton (IC) emission from CR electrons up to TeV energies. 

However, while IC \g-ray emission is dominant, this does not mean that there is no hadron acceleration in \rxj since we see indirect evidence in other SNRs of  efficient hadronic acceleration through other means, e.g. highly amplified magnetic fields and back-reaction of energetic hadrons \citep{parizot06,miceli12}.
 Detailed modeling of CR acceleration coupled to the SNR evolution has been performed by \citet{ellison10} and been fit to X-ray and \g-ray data. 
The lack of thermal X-rays implies a low density \citep[$<0.05$ cm$^{-3}$,][]{ellison10} that suppresses hadronic \g-rays. Using these models with a known value
of the ambient density enables us to provide actual measurements and no longer upper limits on the fraction of kinetic energy that is transferred to accelerated hadrons in \rxj ; a key parameter in understanding the role of SNRs as Galactic CR accelerators.
As no thermal X-ray emission from shocked ISM is seen,
detecting proper motion in X-rays remains the best hope for constraining the ambient  density and the age of this remnant.
In addition, while the SNR is commonly associated with the historical supernova SN 393 \citep{wang97}, this association has been debated by \citet{fesen12}
and shock speed measurements can help to estimate the SNR age.

To obtain a clean measurement of proper motion, a sharp filament is needed.  These types of features are not present in the bright, most studied, 
north-west region of the remnant. By inspecting the full mosaic provided by \xmm \citep[Figure 1 of ][]{acero09}, we found  promising filaments  
in the south-east (SE).  
 The  shock front in the SE region  is the most distant shock from the central compact object (indicating the initial location of the supernova). The shock in this region is probably still evolving in the original cavity carved by the wind of the massive progenitor, enabling us to directly probe the cavity density  and the 
age of the SNR.

\section{Data reduction} \label{sec:data}

\begin{figure*}[t]
 
\includegraphics[bb= 140 42 1340 347 ,clip,width=18cm]{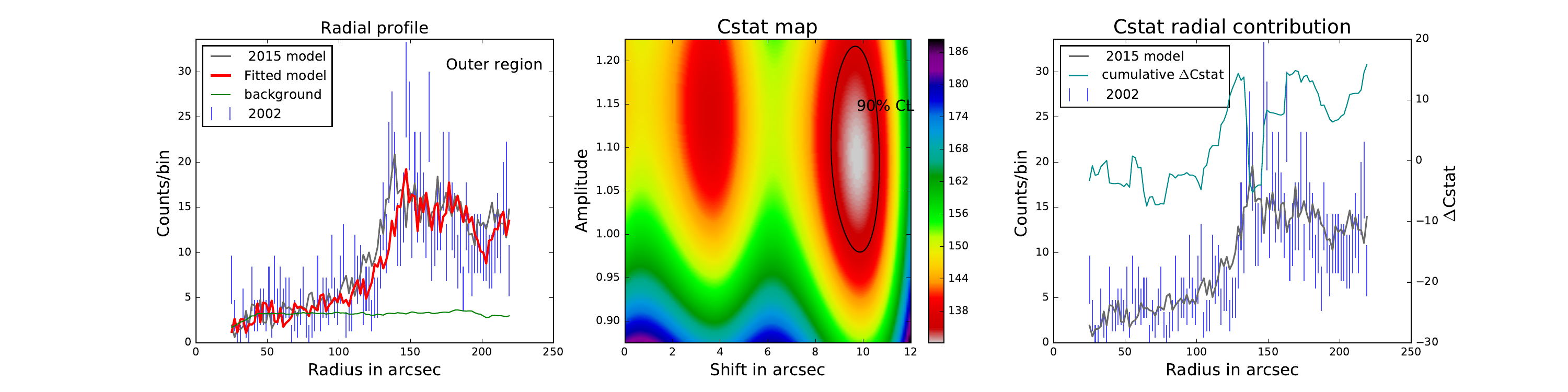}
\includegraphics[bb= 140 14 1340 327 ,clip,width=18cm]{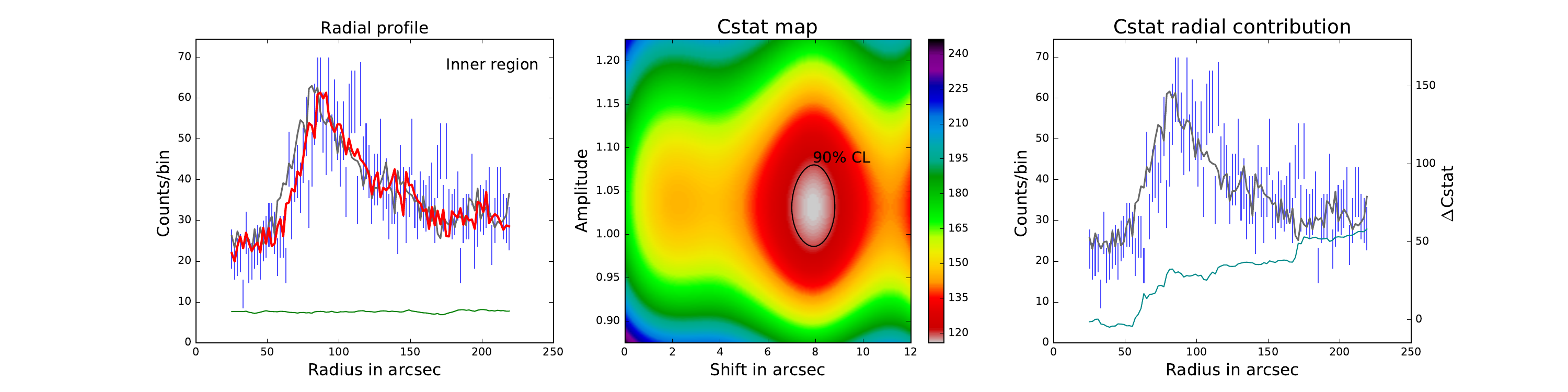}
 
\caption{ Radial profiles, Cstat maps and Cstat radial contribution from the outer region (top panel) and inner region (bottom panel) defined in Figure \ref{XMMimage}.
The right panels show the cumulative statistics radial contribution ($\Delta$Cstat = $C_{\rm 0}(r)-C_{\rm best fit}(r)$) taken at the best-fitted amplitude and where  $C_{\rm 0}(r)$ is the 
Cstat radial contribution at a 0 arcsec model shift. }
\label{profiles}
\end{figure*}

The SE region of the SNR was observed by \xmm on March 14 2002 for 13.7 ks but was  affected by solar flares resulting in a useful exposure time of 11.7 and 5.7 ks for MOS and pn respectively. The deep 82 ks observation (64.1/48.0 ks for MOS and pn respectively after screening) obtained  on March 11 2015 provides high statistics and a 13-year leverage to study the proper motion of the rim. 

 For the image generation and the radial profiles, the instrumental background was derived from the filter wheel closed data provided by the \xmm Science Operation Center \footnote{\url{http://cosmos.esa.int/web/xmm-newton/filter-closed}}.

The 0.6-6 keV energy range was chosen since  there is little emission below 0.6 keV and the instrumental background starts to dominate above 6 keV.
  The background subtracted and exposure corrected MOS + pn image in the 0.6-6 keV energy range (shown in Figure \ref{XMMimage}, left) clearly reveals several sharp filaments.
  The exact shape and edge position of the filaments is not easy to determine by eye, which hampers a proper definition of regions parallel to the shock  to extract radial profiles. To guide our region definition, we used a Canny edge detection algorithm, which is commonly used in image processing. The first step of this algorithm is a Gaussian filtering to reduce the image noise. Then the intensity gradients in the image are computed and potential edges are thinned down to 1-pixel curves. Finally edges pixels are filtered using an hysteresis thresholding on the gradient magnitude.
We used the Canny edge implementation provided in the scikit-image\footnote{\url{http://scikit-image.org}} Python library \citep{skimage} applied to the background subtracted, exposure corrected, count map. In the Gaussian filtering step, the width was set to 20 arcsec radius and the edges are  stable when changing the width of the Gaussian in the filtering step. 
The resulting edges are shown as a contour on Figure \ref{XMMimage} and perfectly outline the shape of the filaments, even in faint regions where the filament fades out.
Based on the edge images shown in Figure \ref{XMMimage} (right), two main structures labeled \textit{outer region} and \textit{inner region} (towards the inside of the SNR which are probably due to projection effects) were identified as promising regions to investigate the filament's proper motion. The inclination of the box regions was chosen to be parallel to the shock edge and the width was chosen so that the shock edge is approximately planar within the regions.

While the 2002 and 2015 observations were carried out with the same position angle, the two pointing positions are separated  by  2.7 arcmin (see crosses in Figure \ref{XMMimage}).
The peak of the sharp inner filament is almost equidistant to the two pointing positions and is not affected by changes in PSFs in 2002 and 2015 (same off-axis angle). 
The outer filament is respectively located at 6.5 arcmin from the 2002 pointing position and 4.3 arcmin from the 2015 pointing position.
We investigated the evolution of the PSF as a function of off-axis angle  using the \xmm calibration file (\textit{XRT*\_XPSF\_0014.CCF}) for a reference energy of 1.5 keV.
The FWHM for our off-axis angle configuration (4.3 arcmin and 6.5 arcmin) corresponds to 5.9 arcsec and 6.3 arcsec respectively.
This very small difference in PSF slightly changes the way the shock front is smoothed but, at the first order, does not affect the localisation of the peak. 

The 2002 and 2015 observations were searched for common registration sources and six candidate X-ray point sources were found  (Sources 1-6 are shown in Figure \ref{XMMimage}, right). Whenever the sources were covered by archival \chandra observations, we  confirmed that they were point-like in nature (Sources 1-3).
Sources 2-4 have optical counterparts and, while sources 3 and 4 have measured proper motions \citep[from the PPMXL catalog,][]{roeser10}, the displacement remains $< 0.5$ arcsec over our 13-year period. Sources 5A (detected in 2002) and 5B (detected in 2015)  are separated by 25 arcsec and are probably two independent variable sources.
We ruled out a source with a high proper motion since source 5B is detected by \chandra in 2000 and 2009 at a position that is compatible with the one obtained by \xmm in 2015.
Source 5A is not detected in the 2000 nor in the 2009 \chandra observation. 
Source 6 has an optical counterpart with high proper motion ($PM_{\rm DEC} = -120$ mas yr$^{-1}$, and  $PM_{\rm RA}$ = 17 mas yr$^{-1}$), which amounts to an 
apparent displacement of 1.6 arcsec over 13 years and is not used in our alignment process. 
Since the 2015 observation is much deeper, we used the 2015 position of sources 1-4 (obtained with the \xmm SAS tool \textit{edetect\_chain}) as reference 
to align the astrometry of the 2002 observation, using the SAS tool \textit{eposcorr}. 
After alignment, the mean 2002-2015 residual offset of the four registration sources is 0.9 arcsec, which we consider as our systematic uncertainty. 

  To avoid any pixelization effect for the rapidly changing profile of the filament, we extracted the source radial profiles with a 2 arcsec binning directly from the event lists and only from the MOS instrument that has a smaller camera pixel size (1.1 arcsec) compared to pn (4.1 arcsec). Removing pn from the 2002 observation (our limiting factor)   only  discards 5.7 ks of observation (11.7 ks of MOS remaining). The instrumental background and the exposure profiles were extracted directly from the filter wheel closed images and the exposure maps. The point sources were removed from the event lists, the exposure maps, and the background maps.   As a result, we obtained three radial profiles (count, exposure, and instrumental background profiles) for each observation and for each MOS camera.

\section{Analysis} \label{sec:analysis}

\begin{figure*}[t]

\includegraphics[bb= 5 5 400 340 ,clip,width=6.1cm]{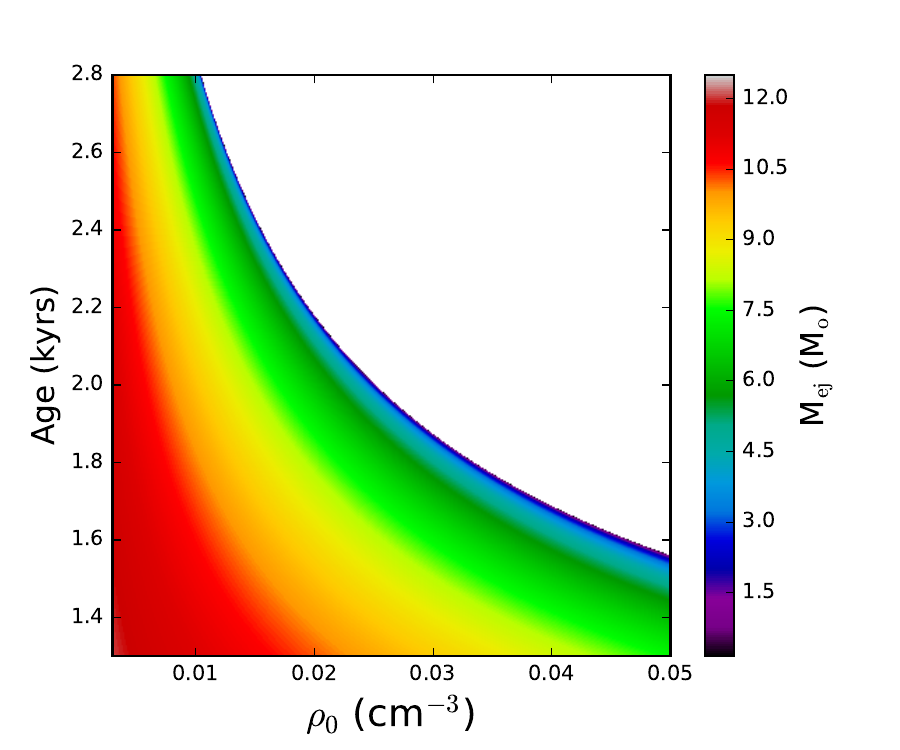}
\includegraphics[bb= 5 5 400 340 ,clip,width=6.1cm]{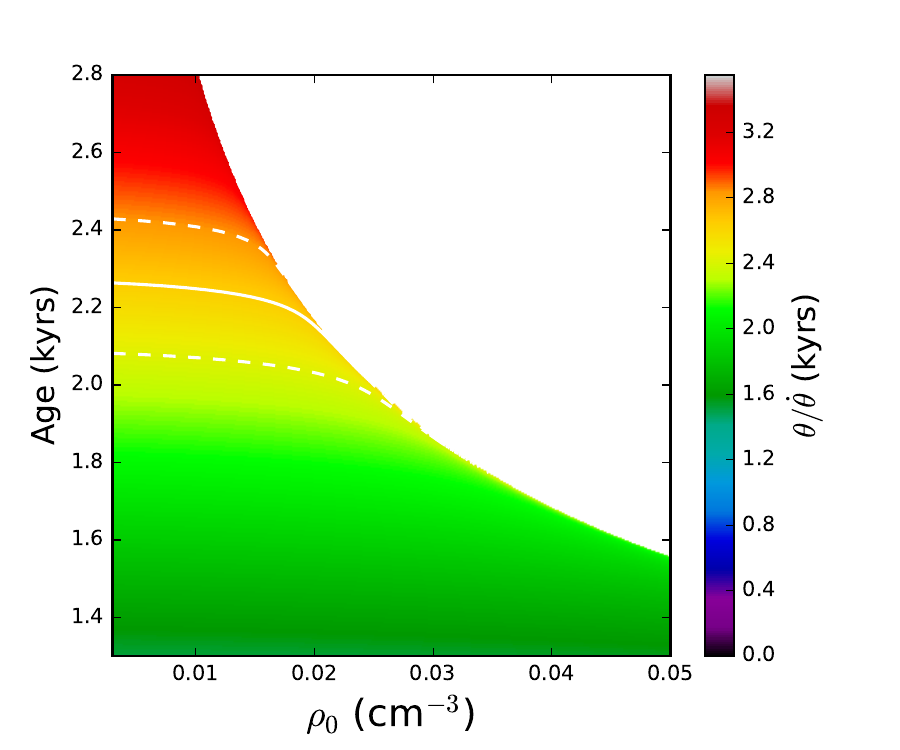}
\includegraphics[bb= 5 5 400 340 ,clip,width=6.1cm]{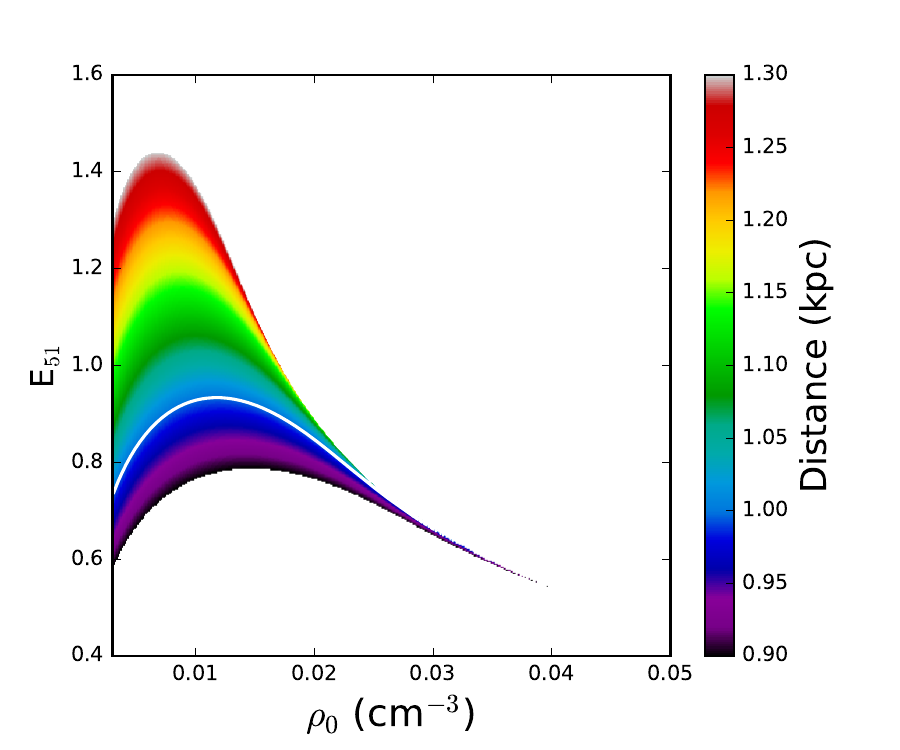}

\caption{
   \textit{Left panel}:  ejecta mass distribution as a function of age and density assuming equality in Eq. \ref{mej}. For each point of the grid, the swept-up mass 
   in the equation is derived by computing the shock radius using evolution equations
   for a supernova explosion energy of 10$^{51}$ ergs, $n=9$, and $s=2$. The white region indicates the excluded parameter space, where the ejecta mass becomes negative.
\textit{Middle panel}: using the ejecta distribution previously derived, the same explosion energy, and ejecta and density profile indices, the ratio $R_{\rm sh}/V_{\rm sh}$ is
obtained and shown here as the distance independent ratio  $\theta/\dot{\theta}$.
The value of this ratio measured in X-rays is shown as a solid white contour  (dashed contours for statistical errors).
 \textit{Right panel}: for a fixed age of 2\,250 yrs derived from the previous panel, the radius of the SNR (or its distance for a fixed angular size) is explored as a function of the
  explosion energy and the wind density. The solid white line represents the radius of the SNR at a distance of 1 kpc. }
\label{age}
\end{figure*}

To estimate the proper motion between the two epochs, we  built a shock profile model based on the 2015 high statistic observation (assuming no errors), 
which is then compared to the 2002 profile. The latter profile is in counts and is the sum of the MOS1 and MOS2 camera. 
We note that the number of counts for some bins of the 2002 radial profile falls below 25 counts/bins and we have therefore 
 used Cash statistics \citep[Cstat,][]{cash79} rather than the $\chi^{2}$ estimator.
A count profile model and the associated Cstat value is calculated as a function of a $Shift$ (in arcsec) and $Amplitude$ (the ratio of the 2002 to 2015 profile) parameters.
The Cstat value $C$ is calculated as

\begin{equation}
C = 2 \sum_{i=1}^{N} M_{\rm i} - S_{\rm i} + S_{\rm i} (ln(S_{\rm i})- ln(M_{\rm i}))
,\end{equation}

where $M_{\rm i}$ is the predicted number of counts in bin $i$ and  $S_{\rm i}$ the observed number of counts in bin $i$ in the 2002 observation.
The model $M$ is calculated for each pair of  parameters and provides the expected number of counts, taking into account the observational conditions of the 2002 observation (exposure and instrumental background profile):

\begin{equation}
M = (Shift(F_{src}^{mod}) * Amp + F_{cst}) * Exp^{obs} + C_{bkg}^{obs}
,\end{equation}

where $F_{src}^{mod}$ represents the vignetting-corrected, background-subtracted flux profile from the 2015 model observation. $F_{cst}$ is the astronomical background flux, 
 measured  upstream of the SNR shock  for $r<75$ arcsec in the outer region (see Figure \ref{profiles}, top, left panel). The parameters $Exp^{obs}$ and $C_{bkg}^{obs}$ represent the 2002 exposure and instrumental background profiles. 
We note that the amplitude and shift parameters are applied only to $F_{src}^{mod}$ not to shift or amplify the instrumental background or any exposure related features.
The resulting 2D Cstat maps are shown in the middle panels of Figure \ref{profiles}, and a 1D profile as a fit of the shift alone is shown in the right panels.
The best-fit results are listed in Table \ref{PMvalues}.
In both regions, no significant change in brightness is observed ($Amplitude$ is compatible with unity) and for a distance of 1 kpc, the corresponding shock speed is respectively 
$2880\pm240_{\rm stat} \pm 330_{\rm syst}$  km s$^{-1}$  and  $3550^{+250}_{-290} \pm 330_{\rm syst}$ km s$^{-1}$  for the inner and outer regions.

\section{Discussion} \label{sec:discussion}

The measurements of the X-ray expansion rate in the SE region can be used to estimate the age of the SNR 
since the shock there is probably still evolving in the cavity carved by the progenitor and has not yet encountered denser material similar to that in the bright north-west region.
The  age $t$ and the radius $R$ can be related via the expansion parameter $m$ as $R \propto t^{m}$, where the age can be derived as $t=mR/V=m \theta/\dot{\theta}$.
The value of the expansion parameter ranges from $m = (n-3)/(n-s)$ in the ejecta dominated
phase to $m = 2/(5-s)$ in the Sedov-Taylor phase, where $n$ and $s$ are the ejecta and ambient medium density profile indices ($\rho(r) \propto r^{-n,s}$).
For the values $n=9$ and $s=2$ (the shock is evolving in the wind of the progenitor star), $m=0.86/0.66$ in the ejecta dominated/Sedov-Taylor phase, respectively. 

In the outer and inner region, we obtain very similar values of $\dot{\theta}/\theta$ (see  Table \ref{PMvalues}) suggesting a projection effect for the inner shock. 
For simplicity, we use the value from the outer shock propagating in the plane of 
the sky, where  $\dot{\theta}/\theta=0.38^{+0.027}_{-0.031} \pm 0.035_{\rm syst}$ kyr$^{-1}$ or $\theta/\dot{\theta}=2.64^{+0.19}_{-0.21} \pm 0.26_{\rm syst}$ kyr.

 With the previously derived values of $\theta$, $\dot{\theta}$, and $m$ and without any assumptions on the ambient density nor on the distance to the SNR,
 we can constrain the age of the SNR to be 1\,760 yrs in the Sedov-Taylor phase and  2\,270 yrs in the ejecta dominated phase (2\,170 yrs if $n=7$).

The characteristic transition timescale between the ejecta dominated and ambient medium dominated phase is given by Eq. 24 of \citet{tang16}. 
For $n=9$ and $s=2$, a reasonable ejecta mass of 10 M$_{\sun}$, an 
 explosion energy of  10$^{51}$ ergs, and an ambient density of 0.05 cm$^{-3}$, $t_{\rm tran}$ is

\begin{equation}
 14\,800 \left( \frac{ M_{\rm ej} } {10 \rm \,M_{\sun}} \right) ^{3/2}   \left( \frac{ \rho_{0}}{0.05\,  \rm cm^{-3}}\right) ^{-1} E_{\rm 51}^{-1/2} \, {\rm yrs}
,\end{equation}

where $\rho_{\rm 0}$ is the density given at a reference radius $R_{\rm 0}=9.6$ pc (the radius of the SE rim at 1 kpc) and $\rho = \rho_{\rm 0} (R/R_{\rm 0})^{-2}$.
Therefore, in the case of \rxj, the SNR is still in the ejecta dominated phase.

To investigate the age and the ambient density parameters in more detail, the shock evolution equations of \citet{tang16},
which provide a better accuracy than the \citet{truelove99} solutions for $s=2$,  are used.
Based on the size of the cavity carved by the progenitor star, \citet{cassam04} estimated a progenitor zero-age main sequence (ZAMS) mass of 12-16 M$_{\sun}$.

The shock in the SE is probably still propagating in the stellar wind since no shock--cloud interaction is detected in $^{12}$CO observations of this region \citep{fukui12}
and the SE rim has the lowest absorption column density \citep{sano15}.
In this scenario, the swept-up mass  $M_{\rm sw}$ in the cavity is related
to the mass loss of the progenitor ($M_{\rm sw} < M_{\rm mass loss}$). Starting from an initial progenitor mass of $M_{\rm zams}=14$ M$_{\sun}$, 
 the  ejecta mass  remaining after  the mass lost in the wind  is
 \begin{equation}
M_{\rm ej} \leq M_{\rm zams} - M_{\rm sw} - M_{\rm cco}
\label{mej}
 ,\end{equation}

    where   for $s=2$, $M_{\rm sw}=4\pi  \rho_{\rm 0} \mu m_{\rm p} R_{\rm 0}^{2} R$ ($\mu$ is the mean atomic mass and $m_{\rm p}$ the proton mass), 
    and $M_{\rm cco}$ is the mass of the central compact object (1.4 M$_{\sun}$). We note that the shock in the south east region is probably not far from the 
cavity boundary (implying $M_{\rm sw} \sim M_{\rm mass loss}$) since the shock in the north west is thought to already be  
interacting with the border of the cavity \citep{fukui12}.

\begin{table*}[ht]

\centering                          
\caption{\label{PMvalues} Best-fitted expansion parameters. }

\begin{tabular}{l c c}        
\hline\hline                 
 Parameter\tablefootmark{a} & Inner region &  Outer region\tablefootmark{d} \\[2pt]
\hline   \\[-5pt]                     
$\Delta\theta$ (arcsec) &   $7.89 \pm 0.67_{\rm stat} \pm 0.9_{\rm syst}$   & $9.73^{+0.70}_{-0.79} \pm 0.9_{\rm syst}$      \\[2pt]
$\dot{\theta}$ (arcsec yr$^{-1}$) &  $0.61\pm0.051_{\rm stat} \pm 0.069_{\rm syst}$  &  $0.75^{+0.05}_{-0.06} \pm 0.069_{\rm syst}$     \\[2pt]
$\theta$ ($^{\circ}$)\tablefootmark{b}&  0.47  &  0.55     \\[2pt]
$\dot{\theta}/\theta$ (kyr$^{-1}$) &  $0.36\pm0.030_{\rm stat} \pm 0.041_{\rm syst}$   &   $0.38^{+0.027}_{-0.031} \pm 0.035_{\rm syst}$  \\[2pt]
$V_{\rm shock}$ (km s$^{-1}$)\tablefootmark{c} &  $2\,880\pm240_{\rm stat} \pm 330_{\rm syst}$   &  $3\,550^{+250}_{-290} \pm 330_{\rm syst}$     \\[2pt]

\hline                                   
\end{tabular}
\tablefoot{
\tablefoottext{a}{Statistical errors are at the 90\% confidence level.}
\tablefoottext{b}{Distance of the rim to the central compact object.}
\tablefoottext{c}{Assuming a distance of 1.0 kpc.}
\tablefoottext{d}{The Cstat profile in the outer region shows a small asymmetry and two-sided errors are reported. }

}
\end{table*}

 For an explosion energy of $E=10^{51}$ ergs,  and  $n=9$ and $s=2$,  the shock radius is computed using evolution equations as a 
 function of the  SNR age and wind density, which is  used to estimate the swept-up mass in the cavity.
The ejecta mass distribution  (Figure \ref{age}, left panel)  is then derived assuming the equality in Eq. \ref{mej}.
 We note that high density values are excluded since the swept-up mass cannot be larger than the progenitor mass ($M_{\rm sw} < M_{\rm zams} - M_{\rm cco}$).

 Using the ejecta mass distribution and the same supernova explosion parameters, $R_{\rm sh}$ and $V_{\rm sh}$ are obtained from evolution equations and expressed independently of the SNR distance as $\theta/\dot{\theta}$ for each age and ambient density shown in Figure \ref{age} (middle panel).
The X-ray measurement of this ratio ($\theta/\dot{\theta}=2.64^{+0.19}_{-0.21} \pm 0.26_{\rm syst}$ kyrs) is shown as a white contour and leads to a SNR age\footnote{The decrease in $\theta/\dot{\theta}$ being slow in Figure  \ref{age} (middle panel), we used the mean value of 2\,250 yrs. } of $2\,250 \pm 170_{\rm stat} \pm 200_{\rm syst}$ yrs.
We note that if the measured proper motion was smaller, the age estimate would increase.

For a fixed age of 2\,250 yrs, the constraints that we can obtain on the wind density depend on the explosion energy and the physical radius of the SNR.
To explore those dependencies, we computed the radius of the SNR (or its distance for a fixed angular size of $\theta = 0.55^{\circ}$) as a function of the  explosion energy and the wind density, as shown in the right-hand panel of Figure \ref{age}.
For a given distance, an upper-limit of the explosion energy can be derived ($E < 10^{51}$ ergs at 1 kpc) since the energy cannot be too high to match the SNR radius.
For a certain given distance and explosion energy, there are two possible density solutions  (a low $\rho$/high $M_{\rm ej}$ and a high $\rho$/low $M_{\rm ej}$ configuration) that can reproduce the SNR radius for a given age.

At a distance of 1 kpc and $E=0.95 \times 10^{51}$ ergs (the maximum value allowed), the corresponding wind density is  $\rho_{\rm 0}=\rho$ = 0.01 cm$^{-3}$ (as $R_{\rm 0}$ was fixed at 9.6 pc ;  the radius at 1 kpc).
For those parameters, the corresponding swept-up mass is 4 M$_{\sun}$ and the ejecta mass is 9 M$_{\sun}$
 and the SNR is therefore still in the ejecta dominated evolution phase. 
The reported mass loss of 4 M$_{\sun}$ is higher than expected for a single progenitor 
with an initial mass of 14 M$_{\sun}$ \citep[about 1-1.5 M$_{\sun}$, see Figure 19 of ][]{sukhbold16}.
If the progenitor evolved in a binary system, the companion could have increased the mass loss by stripping the envelope.
It is, however, not clear if this is the case for \rxj since \citet{katsuda15} have not found the remaining companion.
 
 While the mass loss and ejecta mass are obtained assuming the equality in Eq. \ref{mej}, the ejecta mass could, in fact, be lower since the shock has not fully reached the border of the cavity 
(implying $M_{\rm sw} < M_{\rm mass loss}$) or because the initial progenitor is lower than 14 M$_{\sun}$.
To test this effect, we fixed the ejecta mass to 2 M$_{\sun}$ \citep[as estimated for CasA SNR by][]{hwang12} for all ages and densities.
For this much smaller value, the remnant has just started its transition to the Sedov phase and, by using the same method as shown in Figure \ref{age} (middle panel), an age of $\sim$2\,100 yrs is derived.

 The age of  $2\,250 \pm 170_{\rm stat} \pm 200_{\rm syst}$ yrs derived in the previous paragraph is in tension with the association of \rxj with SN 393. 
 We note that the age of the SNR can be significantly reduced assuming a constant density distribution in the cavity.
 This is plausible for a wind configuration, where the shock would pass little time in the steep ($s=2$) wind profile, and very quickly afterwards evolve 
 in the uniform  shocked wind region ($s=0$).
For $E=10^{51}$ ergs, $d=1$ kpc, $M_{\rm zams} = 14$ M$_{\sun}$, $s=0$ and $n=9$, and $M_{\rm sw}=4/3\pi  \rho \mu m_{\rm p} R{^3}$, 
we obtain $t=1\,760 \pm 130_{\rm stat} \pm 150_{\rm syst}$ yrs  and,  using the same procedure as described in the previous paragraphs, we here obtain two possible density solutions, $\rho=$0.01/0.07 cm$^{-3}$. The higher density solution is however excluded by the upper-limit of 0.02 cm$^{-3}$ based on the lack of thermal X-rays 
in the SE region \citep[region 1 in Figure 8 of ][]{cassam04}.
For $\rho=$0.01 cm$^{-3}$, $M_{\rm sw} = $ 2.5  M$_{\sun}$ for an ejecta mass of 10 M$_{\sun}$  and the SNR is still in the ejecta dominated phase.

For the ejecta index $n=7$ and $M_{\rm zams}$=14 M$_{\sun}$, we obtain an age of $1\,520 \pm 110_{\rm stat} \pm 140_{\rm syst}$ yrs, in agreement with an association with 
SN 393 and a density $\rho=$0.01/0.09 cm$^{-3}$ (the latter being excluded by the lack of thermal X-rays) for a distance of 1 kpc and $E=10^{51}$ ergs.
In this same configuration, it is interesting to note that, if the initial progenitor mass is significantly reduced ($M_{\rm zams}=9$ M$_{\sun}$), 
the maximum explosion energy allowed at 1 kpc is $E=0.8 \times 10^{51}$ ergs (0.6 if d = 0.9 kpc) for a corresponding density of $\rho=$0.02 cm$^{-3}$.
A lower explosion energy can reduce the visual brightness \citep[see, e.g.][]{morozova15} which could help to reconcile the apparent discrepancy noted by \citet{fesen12} between 
an expected visually bright supernova (-3.5 to -5.0 mag) and the historical Chinese record of SN 393 describing a rather faint guest star.

\begin{acknowledgements}
We acknowledge the use of several open-source libraries for Python, including Numpy, Matplotlib, and Astropy.
\end{acknowledgements}


\end{document}